\documentclass[12pt,a4paper]{article}

\usepackage{amsmath,amssymb,amsfonts}
\usepackage{graphicx}
\usepackage[margin=2.5cm]{geometry}
\usepackage[colorlinks=true,citecolor=blue,linkcolor=blue,urlcolor=blue]{hyperref}
\usepackage{physics}
\usepackage{bm}
\usepackage[utf8]{inputenc}
\usepackage[T1]{fontenc}
\usepackage{mathptmx}
\usepackage[numbers,sort&compress]{natbib}
\usepackage{setspace}
\usepackage{caption}
\usepackage[section]{placeins}

\newcommand{\kpar}{k_{\parallel}}
\newcommand{\kperp}{k_{\perp}}
\newcommand{\tpar}{t_{\parallel}}
\newcommand{\tperp}{t_{\perp}}
\newcommand{\vF}{v_{\mathrm{F}}}
\newcommand{\EF}{E_{\mathrm{F}}}
\newcommand{\calP}{\mathcal{P}}

\begin{document}

\title{Parabolic-Cylinder Approach to Valley-Polarized Conductance in Tilted Anisotropic Dirac-Weyl Systems}

\author{Can Yesilyurt$^{1,*}$}
\maketitle

\begin{center}
$^1$Nanoelectronics Research Center, Istanbul, Turkey\\[4pt]
$^*$Corresponding author: can@nanorescenter.com, can--yesilyurt@hotmail.com
\end{center}

\vspace{10pt}

\begin{abstract}
We develop a parabolic-cylinder approach to valley-polarized conductance in tilted anisotropic Dirac-Weyl systems, showing that the smooth-interface scattering problem can be reduced analytically to the Weber equation, which belongs to the same differential-equation class as the quantum harmonic oscillator. This reduction yields closed-form expressions for the angular transmission envelope and clarifies the distinct roles of the tilt components: the perpendicular tilt renormalizes the tunneling envelope width, while the parallel tilt shifts the Fabry-P\'erot resonance structure differently in opposite valleys. Combined with the nonlinear mapping between the fixed device frame and the rotated barrier frame, this analytical structure provides a direct route from valley-dependent interface tunneling to net valley-polarized conductance. We apply the formalism to rotated electrostatic barriers and construct phase diagrams over barrier angle, tilt strength, width, height, and Fermi energy. The results reveal a robust optimum near $t \approx 0.2$ over the parameter range studied, identify the crossover from oscillatory to monotonic polarization regimes, and delineate practical operating windows for candidate materials including 8-$Pmmn$ borophene and WTe$_2$.
\end{abstract}

\section{Introduction}

Valley degrees of freedom in two-dimensional Dirac materials have attracted intense interest as carriers of quantum information, offering an alternative to spin-based electronics~\cite{Rycerz2007,Xiao2007,Schaibley2016,Mak2012,Mak2014valley}. In systems with two or more inequivalent valleys in the Brillouin zone---such as graphene, transition metal dichalcogenides (TMDs), and Weyl semimetals---the valley index $\tau = \pm 1$ is a conserved quantum number that can, in principle, be used for information encoding and processing. A variety of mechanisms have been proposed for generating valley-polarized currents, including strain engineering~\cite{Fujita2010,Pereira2009,Cavalcante2016,Yesilyurt2016graphene,Settnes2016,Milovanovic2014}, line defects~\cite{Gunlycke2011}, magnetic barriers~\cite{Zhai2012,Moldovan2012,Zhai2010}, optical pumping~\cite{Mak2012,ChenJ2022,Wild2023}, chiral anomaly-based transport~\cite{ChenJ2022}, and tilted-cone tunneling asymmetry~\cite{Yesilyurt2019,Nguyen2018,Zhang2023TVHE,Zhang2018borophene,Kong2021}.

Tilted Dirac and Weyl semimetals provide a particularly natural platform for valley-dependent transport. In these materials, the low-energy Hamiltonian contains a tilt term $\hbar v_t \mathbf{k} \cdot \hat{n}$ that breaks the particle-hole symmetry of the Dirac cone, with the tilt direction reversing between opposite valleys ($K$ and $K'$)~\cite{Yesilyurt2017APL,Yesilyurt2017JAP,Yesilyurt2016}. When carriers scatter from an electrostatic barrier, this valley-dependent anisotropy produces distinct transmission profiles $T_K(\theta) \neq T_{K'}(\theta)$ as functions of incidence angle $\theta$~\cite{Yesilyurt2019,Nguyen2018,Zhang2018borophene,Kong2021}. Recently, it has been demonstrated that a single electrostatic barrier rotated by an angle $\phi$ relative to the transport direction produces finite valley-polarized conductance in tilted Dirac systems---without requiring magnetic fields, strain, or optical excitation---using a generalized transfer matrix formalism and semiclassical trajectory simulations~\cite{Yesilyurt2025}. Related work has explored valley filtering in tilted Dirac materials via electrostatic waveguides~\cite{Zheng2020,Ng2022} and tunneling valley Hall currents~\cite{Zhang2023TVHE}.

Previous treatments of valley-dependent tunneling in tilted systems~\cite{Yesilyurt2019,Yesilyurt2025,Nguyen2018} have relied primarily on numerical transfer matrix calculations, which, while accurate, obscure the underlying analytical structure. In this work, the scattering problem at a smooth potential interface in a tilted Dirac system is shown to map exactly onto the \emph{parabolic cylinder equation}---mathematically identical to the quantum harmonic oscillator---with the tilt parameter modifying the effective oscillator quantum number. This mapping provides physical transparency that is not available from the transfer matrix alone: the perpendicular tilt component $\tperp$ renormalizes the oscillator frequency (controlling the tunneling envelope width), while the parallel component $\tpar$ shifts the Fabry-P\'{e}rot resonance positions differently for each valley.

Using this framework, comprehensive phase diagrams are presented that map the valley polarization across the full parameter space of tilt, barrier rotation angle, barrier width, height, and Fermi energy. The analytical perspective reveals several features that are not apparent from individual transfer matrix calculations: (i) a universal polarization peak near $t \approx 0.2$ that arises when the tilt-induced spectral shift becomes commensurate with the Fabry-P\'{e}rot fringe spacing $\sim \pi/d$; (ii) a clear transition from an oscillatory to a monotonic polarization regime as tilt increases; and (iii) the identification of optimal operating windows for candidate tilted Dirac materials, including 8-$Pmmn$ borophene~\cite{LopezBezanilla2016,Zabolotskiy2016} and WTe$_2$~\cite{Wang2016WTe2}.

\section{Model and Analytical Framework}

\subsection{Tilted Dirac Hamiltonian}

We consider a two-dimensional tilted Dirac fermion described by the Hamiltonian
\begin{equation}
  \hat{H} = \hbar \vF (\hat{k}_x \sigma_x + \hat{k}_y \sigma_y)
           + \hbar v_t \hat{k}_y \, \sigma_0 + V(x),
  \label{eq:H_tilted}
\end{equation}
where $\vF$ is the Fermi velocity, $v_t = t \cdot \vF$ is the tilt velocity with dimensionless tilt parameter $t$ ($|t| < 1$ for type-I), $\sigma_{x,y}$ are the Pauli matrices acting on the sublattice pseudospin, $\sigma_0$ is the identity, and $V(x)$ is an electrostatic potential. The tilt is oriented along $y$ (parallel to the barrier for an unrotated barrier along the $y$-axis) and reverses sign between valleys~\cite{Yesilyurt2019,Yesilyurt2017APL}:
\begin{equation}
  v_t \to \tau \, v_t, \qquad \tau = +1 \; (K), \quad \tau = -1 \; (K').
  \label{eq:valley_tilt}
\end{equation}

The energy dispersion of Eq.~\eqref{eq:H_tilted} with $V = 0$ reads
\begin{equation}
  E = \hbar v_t k_y \pm \hbar \vF \sqrt{k_x^2 + k_y^2},
  \label{eq:dispersion}
\end{equation}
describing a Dirac cone tilted along $k_y$. The isoenergy contours are ellipses centered at $k_y = -v_t E / [\vF^2(1 - t^2)]$, with the displacement being valley-dependent through $v_t \to \tau v_t$.

\subsection{Mapping to the Parabolic Cylinder Equation}

Consider a smooth $pn$ junction described by a linear potential $V(x) = eFx$, where $F$ is the electric field. The Dirac equation for the Hamiltonian in Eq.~\eqref{eq:H_tilted}, with transverse momentum $k_y$ conserved, can be written as a coupled system for the two-component spinor $\psi = (\psi_A, \psi_B)^T$. Eliminating one component yields a second-order equation for the other.

Introducing the dimensionless coordinate
\begin{equation}
  \xi = \frac{x - x_0}{\ell}, \qquad
  \ell = \sqrt{\frac{\hbar \vF}{eF}},
  \label{eq:xi_def}
\end{equation}
where $x_0$ is the classical turning point and $\ell$ is the junction length scale, the resulting equation takes the form
\begin{equation}
  \frac{d^2 \varphi}{d\xi^2} + \left( n + \frac{1}{2} - \frac{\xi^2}{4} \right) \varphi = 0.
  \label{eq:weber}
\end{equation}
This is the \emph{parabolic cylinder equation}, also known as the Weber equation. It belongs to the same mathematical family as the quantum harmonic oscillator, although here it appears in the scattering form appropriate to a smooth Dirac interface rather than as a bound-state quantization problem. The effective Weber index is
\begin{equation}
  \boxed{
    n = -\frac{k_y^2 \, \ell^2}{1 - \tperp^2}
  }
  \label{eq:quantum_number}
\end{equation}
where $\tperp$ is the tilt component perpendicular to the interface (along the barrier normal). This is the central result of the harmonic oscillator mapping: the transverse momentum $k_y$ plays the role of the displacement from the oscillator equilibrium, and the tilt parameter renormalizes the effective oscillator frequency through the factor $(1 - \tperp^2)^{-1}$.

The physical interpretation is transparent. For $n = 0$ (normal incidence, $k_y = 0$), the effective potential has zero barrier width---corresponding to Klein tunneling with $T = 1$. For $n < 0$ (oblique incidence), the particle must tunnel through the classically forbidden region of the inverted harmonic oscillator, giving exponentially suppressed transmission. For $n > 0$ (over-the-barrier regime), transmission approaches unity.

\subsection{Transmission from Connection Formulas}

The solutions of Eq.~\eqref{eq:weber} are the parabolic cylinder functions $D_n(\xi)$. Using the asymptotic connection formulas for $D_n(\xi)$ as $\xi \to \pm \infty$~\cite{Cheianov2006}, the exact transmission coefficient through a smooth $pn$ junction is
\begin{equation}
  T(k_y) = \frac{e^{\pi n}}{1 + e^{\pi n}},
  \label{eq:T_pcy}
\end{equation}
which is the Fermi-Dirac function of $\pi n$. In the tunneling regime ($|n| \gg 1$), this reduces to the well-known Gaussian envelope:
\begin{equation}
  T(\theta) \approx \exp\!\left( -\frac{\pi k_\mathrm{F} \, d_\mathrm{eff} \, \sin^2\!\theta}{1 - \tperp^2} \right),
  \label{eq:T_gaussian}
\end{equation}
where $\theta$ is the incidence angle and $d_\mathrm{eff}$ is the effective junction width. Equation~\eqref{eq:T_gaussian} shows that the tilt \emph{broadens} the transmission lobe via the factor $(1 - \tperp^2)^{-1}$: as $\tperp \to 1$ (approaching the type-II limit), the angular window for transmission widens, consistent with the opening of the tilted Dirac cone.

\subsection{Tilt Decomposition for a Rotated Barrier}

When the electrostatic barrier is rotated by angle $\phi$ relative to the $y$-axis (i.e., the barrier normal makes angle $\phi$ with the $x$-axis), the tilt vector decomposes in the barrier frame as
\begin{align}
  \tperp &= \tau \, t \sin\phi, \label{eq:tilt_perp} \\
  \tpar  &= \tau \, t \cos\phi, \label{eq:tilt_par}
\end{align}
where $\tperp$ and $\tpar$ are the components perpendicular (along the barrier normal) and parallel to the barrier, respectively.

The perpendicular component $\tperp$ modifies the harmonic oscillator quantum number via Eq.~\eqref{eq:quantum_number}, altering the tunneling envelope for both valleys. The parallel component $\tpar$, however, breaks the $k_y \leftrightarrow -k_y$ symmetry \emph{differently for each valley}, since the sign of $\tpar$ depends on $\tau$. This is the microscopic origin of valley-dependent transmission.

\section{Valley Polarization Mechanism}

\subsection{Transfer Matrix for a Square Barrier}

For a finite barrier of height $V_0$ and width $d$ (a $pnp$ junction), we employ the transfer matrix method. In the barrier frame, the conserved momentum is $\kpar$, and the perpendicular wavevector $\kperp$ is obtained from the tilted dispersion:
\begin{equation}
  \left(\varepsilon - \tperp \kperp - \tpar \kpar \right)^2 = \kperp^2 + \kpar^2,
  \label{eq:disp_barrier}
\end{equation}
where $\varepsilon = (E - V) / (\hbar \vF)$ is the dimensionless energy measured from the local Dirac point. This yields a quadratic equation for $\kperp$:
\begin{equation}
  (1 - \tperp^2) \, \kperp^2 
  + 2\tperp(\tpar \kpar - \varepsilon) \, \kperp
  + (\tpar \kpar - \varepsilon)^2 - \kpar^2 = 0.
  \label{eq:kperp_quadratic}
\end{equation}
The two solutions $\kperp^{(\pm)}$ correspond to right-moving and left-moving (or evanescent) modes. The spinor ratio for each mode is
\begin{equation}
  \frac{\psi_B}{\psi_A} = \frac{\varepsilon - \tperp \kperp - \tpar \kpar}{\kperp - i \kpar}.
  \label{eq:spinor_ratio}
\end{equation}

The $2 \times 2$ transfer matrix through the barrier region of width $d$ is constructed as $M = S_{\mathrm{out}}^{-1} \, S_{\mathrm{in}} \, P_{\mathrm{in}} \, S_{\mathrm{in}}^{-1} \, S_{\mathrm{out}}$, where $S$ is the spinor mode matrix and $P = \mathrm{diag}(e^{i \kperp^{(+)} d}, \, e^{i \kperp^{(-)} d})$ is the propagation matrix. The transmission probability is
\begin{equation}
  T(\kpar) = \frac{1}{|M_{11}|^2}.
  \label{eq:T_from_M}
\end{equation}

The harmonic oscillator mapping enters here as the \emph{envelope}: the PCY result of Eq.~\eqref{eq:T_pcy} provides the smooth tunneling envelope that modulates the Fabry-P\'{e}rot oscillations arising from multiple reflections inside the barrier. The Fabry-P\'{e}rot resonance condition is
\begin{equation}
  \kperp^{(+)} \, d = m\pi, \quad m = 0, 1, 2, \ldots
  \label{eq:FP_condition}
\end{equation}
Since $\kperp^{(+)}$ depends on $\tpar$ (and hence on the valley index $\tau$), the resonance positions in $\kpar$-space are valley-dependent, producing $T_K(\kpar) \neq T_{K'}(\kpar)$.

\subsection{The Nonlinear Device-to-Barrier Frame Mapping}

A crucial aspect of the valley polarization mechanism, which has not been emphasized in previous work~\cite{Nguyen2018,Zhang2023TVHE}, is the role of the coordinate transformation between the fixed device frame $(x,y)$ and the rotated barrier frame $(\parallel,\perp)$. We distinguish the fixed device frame $(x,y)$, where transport is measured and the conserved transverse momentum is labeled by $k_y$, from the rotated barrier frame $(\parallel,\perp)$ obtained by rotating the axes by the barrier angle $\phi$. In the device frame, an incoming carrier from the left lead has momentum $(k_x, k_y)$ with $k_x = +\sqrt{k_\mathrm{F}^2 - k_y^2}$ (right-moving). When the barrier normal is rotated by angle $\phi$, the conserved barrier-frame momentum is
\begin{equation}
  \boxed{
    \kpar(k_y) = -\sqrt{k_\mathrm{F}^2 - k_y^2} \, \sin\phi + k_y \cos\phi
  }
  \label{eq:nonlinear_mapping}
\end{equation}
This mapping is \emph{nonlinear} in $k_y$ due to the square root. Critically, it is \emph{not antisymmetric}:
\begin{equation}
  \kpar(k_y) + \kpar(-k_y) = -2\sqrt{k_\mathrm{F}^2 - k_y^2} \, \sin\phi \neq 0 \quad \text{for } \phi \neq 0.
  \label{eq:asymmetry}
\end{equation}

This broken antisymmetry is the key to understanding why the integrated conductance yields $G_K \neq G_{K'}$. In the barrier frame, time-reversal symmetry guarantees $T_K(\kpar) = T_{K'}(-\kpar)$, so that $\int T_K(\kpar) \, d\kpar = \int T_{K'}(\kpar) \, d\kpar$ identically. However, the conductance is computed by integrating over the device-frame transverse momentum $k_y$:
\begin{equation}
  G_\tau = \frac{e^2}{h} \cdot \frac{W}{2\pi} \int_{-k_\mathrm{F}}^{k_\mathrm{F}} T_\tau\!\big(\kpar(k_y)\big) \, dk_y,
  \label{eq:Landauer}
\end{equation}
where $W$ is the sample width. The nonlinear mapping $\kpar(k_y)$ ensures that the integration measure $dk_y$ samples different regions of the $T_K(\kpar)$ and $T_{K'}(\kpar)$ landscapes, breaking the cancellation guaranteed in the barrier frame.

Physically, this can be understood as follows: the rotated barrier acts as a prism that refracts carriers differently for the two valleys. The tilt shifts the Fabry-P\'{e}rot resonances in $\kpar$-space, while the nonlinear mapping determines \emph{which} device-frame angles map onto these resonances. The combination produces a net valley-polarized current even though time-reversal symmetry holds microscopically.

\subsection{Valley Polarization}

The valley polarization is defined as
\begin{equation}
  \calP = \frac{G_K - G_{K'}}{G_K + G_{K'}}.
  \label{eq:polarization}
\end{equation}
From the analysis above, we can identify two necessary conditions for $\calP \neq 0$:
\begin{enumerate}
  \item \textbf{Nonzero tilt} ($t \neq 0$): Required to produce valley-dependent scattering, $T_K \neq T_{K'}$.
  \item \textbf{Nonzero barrier rotation} ($\phi \neq 0$): Required to break the antisymmetry of the lab-to-barrier mapping, converting the angle-resolved valley contrast into net current polarization.
\end{enumerate}

At $\phi = 0$, the mapping $\kpar(k_y) = k_y$ is linear and antisymmetric, so $G_K = G_{K'}$ exactly regardless of tilt. At $\phi = 90^\circ$, the tilt is entirely perpendicular to the barrier ($\tpar = 0$), so $T_K = T_{K'}$ and no valley contrast exists. The optimal polarization occurs at intermediate angles where both the tilt-induced spectral asymmetry and the nonlinear coordinate warping are significant.

This symmetry structure is worth emphasizing because it cleanly separates the two layers of the problem. The smooth-interface tunneling physics is controlled analytically by the Weber-class envelope, whereas the measurable valley-polarized conductance emerges only after the rotated geometry projects that valley contrast through the nonlinear laboratory-frame integration. In other words, the exact solvable local scattering problem and the device-level valley filtering are distinct ingredients, and both are required.

\section{Results and Discussion}

We now present numerical results computed using the analytical transfer matrix framework developed above. Unless stated otherwise, the default parameters are: Fermi energy $\EF = 100$~meV, barrier height $V_0 = 200$~meV, barrier width $d = 50$~nm, and sample width $W = 1$~$\mu$m. The Fermi velocity is $\vF = 10^6$~m/s, typical of Dirac materials.

\subsection{Harmonic Oscillator Mapping and Valley Splitting}

Figure~1 demonstrates the core analytical framework. Panel~(a) shows the nonlinear mapping $\kpar(k_y)$ of Eq.~\eqref{eq:nonlinear_mapping} for several barrier rotation angles. At $\phi = 0$, the mapping is the identity ($\kpar = k_y$), while increasing $\phi$ introduces a progressively nonlinear, downward-bent curve. This bending is what breaks the antisymmetry and enables net valley polarization. Panel~(b) shows the PCY transmission envelope from Eq.~\eqref{eq:T_pcy} for different tilt values, illustrating the broadening of the tunneling lobe with increasing $\tperp$ through the $(1 - \tperp^2)^{-1}$ factor.

Panels~(d)--(f) show the central result: valley-resolved transmission $T(\theta)$ in the device frame for $t = 0.5$ at barrier rotation angles $\phi = 0^\circ$, $15^\circ$, and $25^\circ$. At $\phi = 0^\circ$ [panel~(d)], the two valley profiles are perfect mirrors about $\theta = 0$, yielding $\calP = 0$. The Fabry-P\'{e}rot resonances are clearly visible as sharp transmission peaks within the PCY envelope. As $\phi$ increases to $15^\circ$ [panel~(e)], the valley profiles become asymmetric in the device frame due to the nonlinear mapping, yielding $\calP = 0.58$. At $\phi = 25^\circ$ [panel~(f)], valley $K'$ is almost completely filtered out over the accessible angular range, giving $\calP = 0.95$. The mechanism is clear: the nonlinear mapping concentrates valley $K$ transmission into the forward-scattering window while pushing valley $K'$ resonances to large (inaccessible) angles.

\begin{figure*}[!htb]
  \centering
  \includegraphics[width=0.88\textwidth]{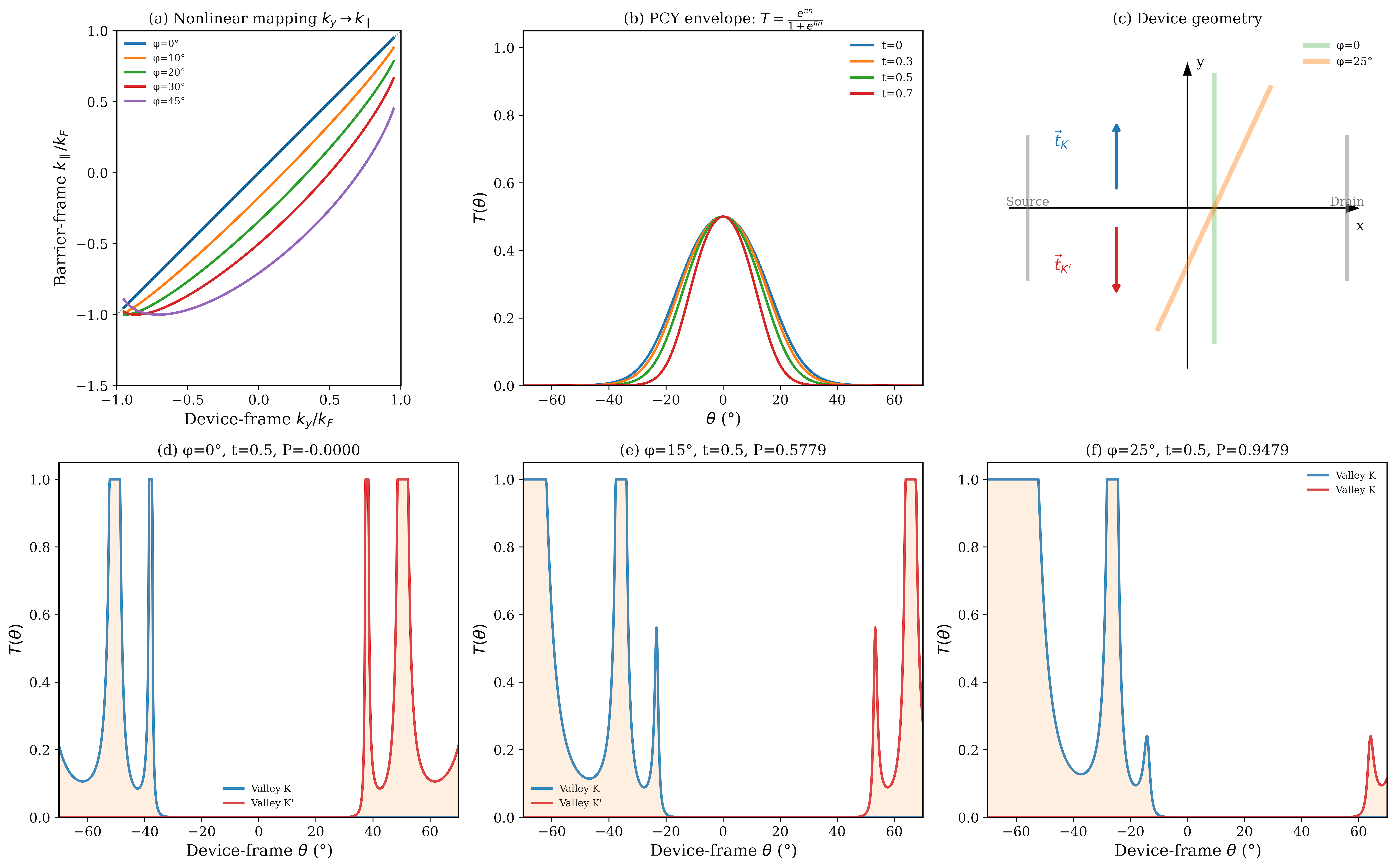}
  \caption{\textbf{Harmonic oscillator mapping and valley-resolved transmission.}
  (a)~Nonlinear mapping from device-frame transverse momentum $k_y$ to barrier-frame conserved momentum $\kpar$ [Eq.~\eqref{eq:nonlinear_mapping}] for several barrier rotation angles $\phi$. At $\phi = 0$, the mapping is linear ($\kpar = k_y$); increasing $\phi$ introduces asymmetry that breaks time-reversal-protected valley degeneracy in the integrated conductance.
  (b)~Parabolic cylinder function (PCY) transmission envelope [Eq.~\eqref{eq:T_pcy}] for a smooth $pn$ junction at different tilt parameters $t$. The tilt broadens the angular transmission window through the factor $(1 - t^2)^{-1}$ in the harmonic oscillator quantum number.
  (c)~Device geometry: transport along $x$, electrostatic barrier along $y$ (green), tilt directed along $\pm y$ for valleys $K$/$K'$ (blue/red arrows), with a rotated barrier at angle $\phi$ (orange).
  (d--f)~Valley-resolved transmission $T(\theta)$ in the device frame for tilt $t = 0.5$ at barrier rotation angles $\phi = 0^\circ$, $15^\circ$, and $25^\circ$. At $\phi = 0^\circ$, the valley profiles are mirror-symmetric and $\calP = 0$. Increasing $\phi$ breaks this symmetry through the nonlinear coordinate mapping, yielding $\calP = 0.58$ at $\phi = 15^\circ$ and $\calP = 0.95$ at $\phi = 25^\circ$. Parameters: $\EF = 100$~meV, $V_0 = 200$~meV, $d = 50$~nm.}
  \label{fig:core_physics}
\end{figure*}
\FloatBarrier

Figure~2 presents comprehensive parameter sweeps of the valley polarization. Panel~(a) shows $\calP$ versus barrier rotation angle for different tilt values. For all tilts, $\calP = 0$ at $\phi = 0$ (confirming the symmetry argument of Eq.~\eqref{eq:asymmetry}) and increases monotonically with $\phi$, approaching unity at moderate angles. The rate of increase is faster for larger tilt values, as expected from the larger valley splitting $\Delta \kpar \propto t$. The curves for $t = 0.2$--$0.6$ show fine oscillations at small $\phi$, arising from Fabry-P\'{e}rot resonances cycling in and out of the integration window.

Panel~(b) reveals a notable feature: all barrier angles exhibit a peak in polarization near $t \approx 0.2$. This appears to be a universal feature arising from the interplay between the elliptical isoenergy contour geometry and the device geometry. At very small tilt, the valley splitting is negligible. At large tilt, the cone opens so broadly that both valleys transmit over wide angular ranges, reducing the relative asymmetry. The optimal tilt $t^* \approx 0.2$ corresponds to the condition where the tilt-induced $\kpar$ shift becomes comparable to the Fabry-P\'{e}rot fringe spacing $\Delta \kpar^{\mathrm{FP}} \sim \pi/d$.

Panel~(c) shows the individual valley conductances $G_K$ and $G_{K'}$ versus $\phi$ for $t = 0.5$. Valley $K$ conductance increases with $\phi$ while $G_{K'}$ is suppressed, confirming that the barrier rotation acts as a valley-selective filter. The valley conductance curves obtained from the harmonic oscillator / transfer matrix approach reproduce the results of the independent transfer matrix study in Ref.~\cite{Yesilyurt2025}, providing a cross-validation of the two complementary formalisms. Since the asymptotic regions on the left and right of the barrier are identical in the present geometry, the transmission expression in Eq.~\eqref{eq:T_from_M} can be used without an additional left-right velocity prefactor.

Panel~(d) shows the polarization as a function of barrier width for $t = 0.5$, $\phi = 20^\circ$. The oscillatory behavior is a direct signature of Fabry-P\'{e}rot resonances: as $d$ increases, the resonance condition $\kperp d = m\pi$ cycles through constructive and destructive interference conditions differently for each valley. The overall trend is an increase of $\calP$ with $d$, since wider barriers are more selective.

Panel~(e) displays $\calP$ versus barrier height $V_0$. The polarization is near zero for $V_0 < \EF = 100$~meV, since in this regime the barrier does not support a hole pocket and transmission is non-resonant. Above $V_0 \approx 2\EF$, $\calP$ saturates near unity.

Panel~(f) shows $\calP$ versus Fermi energy with a peak near $\EF \approx V_0/2 = 100$~meV. We note, however, that at this energy the total conductance is suppressed (since the Fermi level lies near the barrier top), so the high polarization at $\EF = V_0/2$ may not be practically useful. A device should be operated at $\EF$ values where both high polarization and appreciable total conductance are achieved simultaneously.

\begin{figure*}[!htb]
  \centering
  \includegraphics[width=0.88\textwidth]{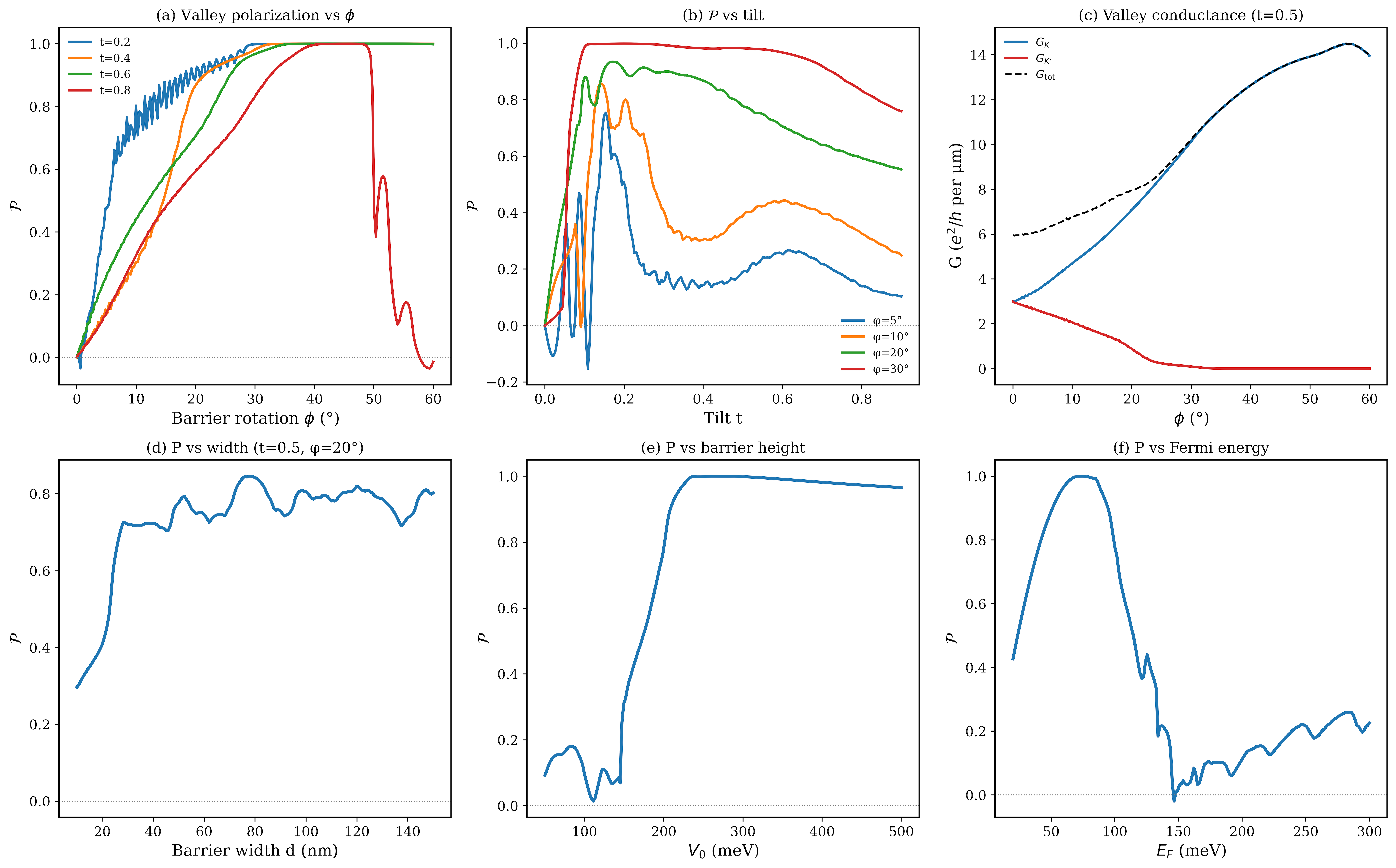}
  \caption{\textbf{Valley polarization parameter dependences.}
  (a)~$\calP$ versus barrier rotation angle $\phi$ for tilt values $t = 0.2$, $0.4$, $0.6$, and $0.8$. All curves start at $\calP = 0$ for $\phi = 0$ (confirming the symmetry argument) and increase monotonically, with fine Fabry-P\'{e}rot oscillations visible at small $\phi$ for moderate tilts.
  (b)~$\calP$ versus tilt parameter $t$ for fixed barrier angles $\phi = 5^\circ$, $10^\circ$, $20^\circ$, and $30^\circ$. A common polarization peak appears near $t \approx 0.2$ for all angles, corresponding to the condition where the tilt-induced $\kpar$ shift matches the Fabry-P\'{e}rot fringe spacing $\sim\!\pi/d$.
  (c)~Individual valley conductances $G_K$ and $G_{K'}$ versus $\phi$ for $t = 0.5$, showing the valley-selective filtering: $G_K$ increases while $G_{K'}$ is progressively suppressed with barrier rotation.
  (d)~$\calP$ versus barrier width $d$ for $t = 0.5$, $\phi = 20^\circ$. The oscillatory behavior reflects Fabry-P\'{e}rot resonances cycling through constructive/destructive conditions differently for each valley, with a general upward trend as wider barriers are more selective.
  (e)~$\calP$ versus barrier height $V_0$. Polarization vanishes for $V_0 < \EF$ (no hole pocket inside the barrier) and saturates near unity for $V_0 \gtrsim 2\EF$.
  (f)~$\calP$ versus Fermi energy $\EF$ for $t = 0.5$, $\phi = 20^\circ$, $V_0 = 200$~meV. The peak near $\EF \approx V_0/2$ coincides with low total conductance; practical operation requires $\EF$ values offering both high $\calP$ and appreciable $G_{\mathrm{tot}}$.}
  \label{fig:parameter_sweeps}
\end{figure*}
\FloatBarrier

Figure~3 provides a global view of the valley filtering landscape. Panel~(a) shows the device-frame $T(\theta)$ for $t = 0.5$, $\phi = 20^\circ$, highlighting the dramatic angular separation of valley $K$ and $K'$ transmission profiles. Valley $K$ dominates the negative-$\theta$ hemisphere with prominent Fabry-P\'{e}rot resonances, while $K'$ is confined to a narrow peak near $\theta \approx +60^\circ$.

Panel~(b) maps the valley contrast $\Delta T = T_K - T_{K'}$ as a function of incidence angle $\theta$ and barrier rotation $\phi$. The diagonal structure reflects the systematic shift of the transmission profiles with $\phi$ due to the nonlinear mapping. The antisymmetric pattern about $\theta = 0$ at $\phi = 0$ progressively breaks as $\phi$ increases, with the $T_K > T_{K'}$ region (red) expanding to dominate the angular space.

Panel~(c) presents the phase diagram $\calP(t, \phi)$, which is the central result of this work. Three distinct regimes are apparent: (i)~a low-polarization region at small $t$ and $\phi$ where the valley splitting is insufficient; (ii)~an oscillatory regime at low tilt ($t \lesssim 0.2$) where $\calP$ alternates sign due to individual Fabry-P\'{e}rot fringes falling on alternating valleys; and (iii)~a high-polarization regime ($\calP > 0.8$) occupying a broad region of parameter space at moderate-to-large tilt and rotation angles. The boundary between the oscillatory and monotonic regimes occurs near $t \sim 0.2$, consistent with the universal peak observed in Fig.~2(b).

The phase diagram demonstrates that robust, high valley polarization is achievable over a wide range of experimentally relevant parameters. For materials with moderate tilt ($t = 0.3$--$0.5$), such as 8-$Pmmn$ borophene or certain transition metal ditellurides, a barrier rotation of $\phi \gtrsim 20^\circ$ is sufficient to achieve $\calP > 90\%$.

\begin{figure*}[!htb]
  \centering
  \includegraphics[width=0.88\textwidth]{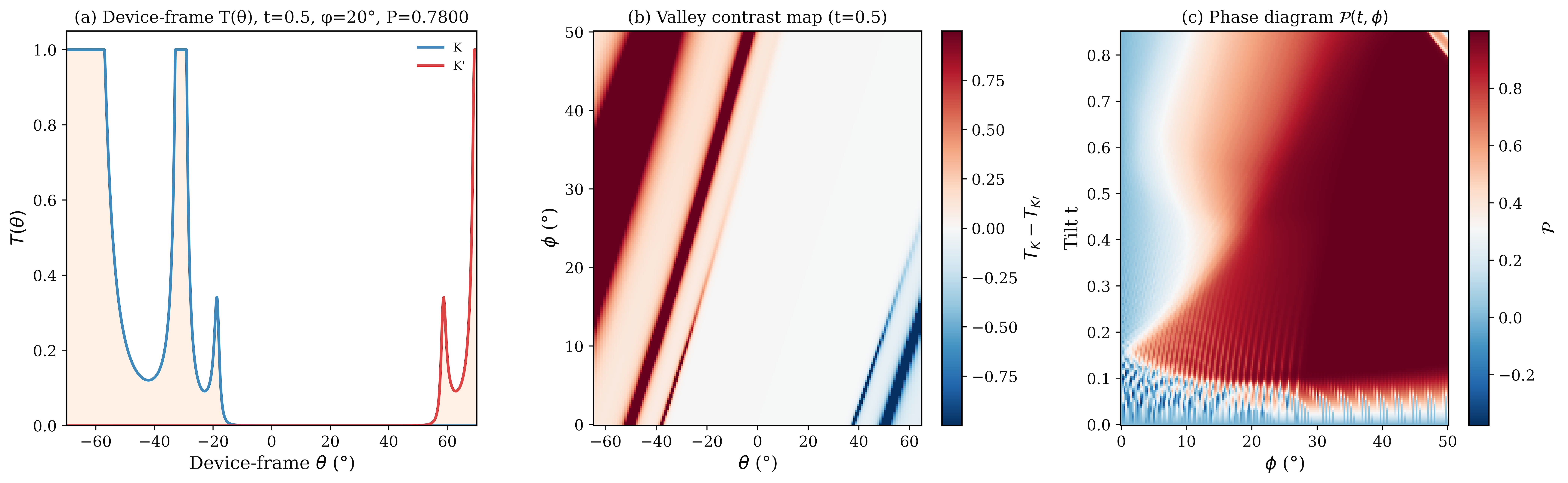}
  \caption{\textbf{Valley contrast map and phase diagram.}
  (a)~Device-frame transmission $T(\theta)$ for valleys $K$ (blue) and $K'$ (red) at $t = 0.5$, $\phi = 20^\circ$, yielding $\calP = 0.78$. Valley $K$ dominates the negative-$\theta$ hemisphere with prominent Fabry-P\'{e}rot resonances, while $K'$ is confined to a narrow peak near $\theta \approx +60^\circ$.
  (b)~Valley contrast $\Delta T = T_K - T_{K'}$ as a function of device-frame incidence angle $\theta$ and barrier rotation $\phi$ for $t = 0.5$. The diagonal structure reflects the systematic shift of valley transmission profiles with $\phi$. At $\phi = 0$, the pattern is antisymmetric about $\theta = 0$ (zero net polarization); increasing $\phi$ breaks this antisymmetry progressively.
  (c)~Phase diagram of valley polarization $\calP(t, \phi)$. Three regimes are evident: (i)~negligible polarization at small $t$ and $\phi$; (ii)~an oscillatory regime at low tilt ($t \lesssim 0.2$) where $\calP$ alternates sign as individual Fabry-P\'{e}rot fringes shift between valleys; and (iii)~a robust high-polarization regime ($\calP > 0.8$) at moderate-to-large $t$ and $\phi$. The transition near $t \sim 0.2$ marks where the tilt-induced $\kpar$ shift exceeds the Fabry-P\'{e}rot fringe spacing. Parameters: $\EF = 100$~meV, $V_0 = 200$~meV, $d = 50$~nm.}
  \label{fig:phase_diagram}
\end{figure*}
\FloatBarrier

\section{Material Candidates and Experimental Feasibility}

The harmonic oscillator framework developed above applies to any two-dimensional material hosting tilted Dirac or Weyl cones with valley-dependent tilt orientations. The phase diagrams in Figs.~2 and~3 can be read directly as design charts for specific materials, once the tilt parameter $t$ and Fermi velocity $\vF$ are known. We now assess the feasibility for three candidate material platforms.

\textbf{8-\textit{Pmmn} borophene.}  First-principles calculations predict that monolayer 8-$Pmmn$ borophene hosts a pair of tilted type-I Dirac cones with a tilt parameter $t \approx 0.4$--$0.5$ and a Fermi velocity $\vF \approx 0.86 \times 10^6$~m/s~\cite{LopezBezanilla2016,Zabolotskiy2016}. The tilt is oriented along $\Gamma$--$Y$, and the two valleys at $(0, \pm k_D)$ exhibit opposite tilt directions, precisely the configuration assumed in the present model. The phase diagram (Fig.~3c) shows that $t = 0.5$ lies squarely within the monotonic, high-polarization regime, where the harmonic oscillator analysis predicts robust valley filtering at barrier rotation angles $\phi \gtrsim 20^\circ$. The Dirac fermion nature of 8-$Pmmn$ borophene has been confirmed experimentally by the observation of characteristic Dirac cone features~\cite{Feng2017}. Furthermore, the tilt can in principle be tuned by atomic substitutions (e.g., carbon for boron)~\cite{Yekta2023}, which would allow traversal of the phase diagram along the tilt axis---a direct experimental test of the oscillatory-to-monotonic transition predicted near $t \approx 0.2$. A key challenge remains the large-area synthesis of this specific borophene polymorph on suitable substrates; current synthesis routes yield $\chi_3$ and $\beta_{12}$ phases on silver substrates, and achieving the 8-$Pmmn$ phase requires further development of growth protocols.

\textbf{WTe$_2$ and related transition metal ditellurides.}  Monolayer and few-layer WTe$_2$ exhibit an anisotropic Dirac-like dispersion with tilted cones~\cite{Wang2016WTe2}, and gate-tunable transport has been demonstrated experimentally at room temperature. The anisotropy axis provides a natural distinction between transport directions. The mechanism requires only that the tilt reverses sign between valleys---a condition satisfied generically in systems with two valleys related by time reversal. The moderate tilt parameter ($t \sim 0.3$--$0.5$ depending on the specific compound and stacking) places WTe$_2$ in the regime where the harmonic oscillator analysis predicts a smooth polarization response without the oscillatory fine structure that complicates operation at low tilt. Several groups have reported high-mobility transport in exfoliated WTe$_2$ devices~\cite{Wang2016WTe2}, and the required rotated gate geometry is readily accessible using standard electron-beam lithography.

\textbf{Organic Dirac materials.}  The layered organic conductor $\alpha$-(BEDT-TTF)$_2$I$_3$ hosts tilted Dirac cones under pressure, with a tilt parameter that can be tuned continuously by adjusting the applied hydrostatic pressure~\cite{Katayama2006}. While the lower Fermi velocity ($\vF \sim 10^4$--$10^5$~m/s) requires proportionally wider barriers or larger gate voltages, the continuously tunable tilt provides a unique platform for systematically exploring the $\calP(t, \phi)$ phase diagram predicted by the harmonic oscillator framework.

\textbf{Device geometry considerations.}  The required device structure consists of a two-terminal geometry (source and drain leads) with a top-gate electrode patterned at an angle $\phi$ relative to the source-drain axis, as described in Ref.~\cite{Yesilyurt2025}. The gate defines the electrostatic barrier whose height $V_0$ is controlled by the gate voltage, and the barrier width $d$ is set by the gate geometry (i.e., the projected width of the gate along the transport direction). For the default parameters ($\EF = 100$~meV, $V_0 = 200$~meV, $d = 50$~nm), the device dimensions are well within the reach of electron-beam lithography, and the required gate voltages ($V_g \sim V_0/e \sim 0.2$~V for a thin dielectric) are compatible with standard CMOS technology. The total conductance remains appreciable throughout the high-polarization regime (Fig.~2c), confirming that the valley filter does not operate by simply suppressing all transport.

\section{Conclusion}

We have developed an analytical framework for valley-resolved quantum transport in tilted Dirac/Weyl semimetals based on the mapping of the scattering problem onto the parabolic cylinder equation---the quantum harmonic oscillator. This approach reveals the physical structure underlying valley-dependent tunneling: the harmonic oscillator quantum number $n = -k_y^2 \ell^2 / (1 - t_\perp^2)$ controls the tunneling envelope, while the valley-dependent parallel tilt $\tpar = \tau t \cos\phi$ shifts the Fabry-P\'{e}rot resonance positions. The net valley polarization in conductance emerges from the nonlinear mapping between the laboratory and barrier reference frames, which breaks the time-reversal-protected cancellation of valley-resolved currents.

Applied to the rotated-barrier geometry established in Ref.~\cite{Yesilyurt2025}, the harmonic oscillator framework provides a transparent analytical basis for results that were previously accessible only through numerical transfer matrix calculations. The comprehensive phase diagrams presented here reveal universal features of the valley polarization landscape---including the characteristic peak at $t \approx 0.2$ governed by the commensurability condition between tilt-induced spectral shifts and Fabry-P\'{e}rot fringe spacing, and the sharp transition from oscillatory to monotonic polarization behavior---that serve as design guidelines for device optimization. We have identified 8-$Pmmn$ borophene and WTe$_2$ as promising candidate platforms, with the continuously tunable tilt in organic Dirac conductors offering a pathway for systematic experimental exploration of the predicted phase diagram.

The harmonic oscillator framework developed here may pave the way for extensions to multi-barrier structures (i.e., cascaded valley filters), smooth potential profiles that naturally match the PCY envelope, and type-II tilted systems, where the over-the-barrier regime ($n > 0$) becomes accessible for certain angular sectors. More broadly, the mapping establishes a connection between valley-resolved transport and the well-studied physics of the quantum harmonic oscillator, suggesting that techniques from quantum optics and strong-field QED (e.g., Landau-Zener transitions, Schwinger pair production) may find analogues in the valley transport of tilted Dirac materials.

\section*{Acknowledgments}

\noindent [To be added.]

\end{document}